\pdfoutput=1
\documentclass{aa}

\usepackage{graphicx}
\usepackage{txfonts}

\usepackage{xcolor}
\usepackage[pdfpagelabels=false]{hyperref}
\hypersetup{colorlinks=true,linkcolor=blue,citecolor=blue,filecolor=blue,urlcolor=blue}
\usepackage{multirow}   
\usepackage{booktabs}   
\usepackage{capt-of}    

\begin{document}

   \title{The interplay between forming planets and photoevaporating discs}

   \subtitle{II: Wind-driven gas redistribution}

   \author{M. L. Weber\inst{1,2}\fnmsep\thanks{ \href{mailto:mweber@usm.lmu.de}{mweber@usm.lmu.de}}
          \and
          G. Picogna\inst{1}
          \and
          B. Ercolano\inst{1,2}
          }

   \institute{University Observatory, Faculty of Physics, Ludwig-Maximilians-Universität München, Scheinerstr. 1, 81679 Munich, Germany
         \and
             Excellence Cluster ORIGINS, Boltzmannstr. 2, 85748 Garching, Germany
             }

   \date{Received <date> / Accepted <date>}

\abstract{
Disc winds and planet-disc interactions are two crucial mechanisms that define the structure, evolution and dispersal of protoplanetary discs. While winds are capable of removing material from discs, eventually leading to their dispersal, massive planets can shape their disc by creating sub-structures such as gaps and spiral arms. 
}
{
We study the interplay between an X-ray photoevaporative disc wind and the substructures generated due to planet-disc interactions to determine how their mutual interactions affect the disc's and the planet's evolution.
}
{
We perform three-dimensional hydrodynamic simulations of viscous ($\alpha = 6.9\cdot10^{-4}$) discs that host a Jupiter-like planet and undergo X-ray photoevaporation. We trace the gas flows within the disc and wind and measure the accretion rate onto the planet, as well as the gravitational torque that is acting on it. 
}
{
Our results show that the planetary gap takes away the wind's pressure support, allowing wind material to fall back into the gap. This opens new pathways for material from the inner disc (and part of the outer disc) to be redistributed through the wind towards the gap. Consequently, the gap becomes shallower, and the flow of mass across the gap in both directions is significantly increased, as well as the planet's mass-accretion rate (by factors $\approx 5$ and $\approx 2$, respectively). Moreover, the wind-driven redistribution results in a denser inner disc and less dense outer disc, which, combined with the recycling of a significant portion of the inner wind, leads to longer lifetimes of the inner disc, contrary to the expectation in a planet-induced photoevaporation (PIPE) scenario that has been proposed in the past.
}
{
}


   \maketitle
%

\section{Introduction}\label{sec:introduction}
The extensive diversity of exoplanets and exoplanetary systems discovered in recent years highlights the importance of understanding the many different processes involved in the formation and evolution of planets in protoplanetary discs.
Two of the most important processes that can affect the structure, evolution, and dispersal of protoplanetary discs are planet-disc interactions and disc winds.

Through the gravitational influence on its surrounding material, a newly formed planet that is still embedded in a protoplanetary disc can generate complex substructures in the disc that can influence both the planet's and the disc's evolution \citep[see e.g.][for a recent review]{paardekooper2022}. 
A planet excites spiral density waves at Lindblad resonances, and if the planet is massive enough, the torque that the spiral waves exert on the disc can become strong enough to overcome the viscous torque, at which point a gap will open in the disc along the planet's orbit \citep{goldreich1979, goldreich1980, papaloizou1984}.
Numerical simulations show that at the edges of such gaps, the gas follows meridional flows, circulating between the lower and upper layers of the protoplanetary disc \citep[e.g.][]{morbidelli2014, fung2016}.
The disc, on the other hand, also has a gravitational influence on the planet and asymmetric substructures in the disc can result in a net torque that acts on the planet, changing its angular momentum, which will lead to its migration through the disc, either radially inward or outward \citep[e.g.][and references therein]{kley2012}.
Recent observations, e.g. with the Atacama Large Millimeter/submillimeter Array (ALMA), have revealed that most discs are indeed highly substructured, including, among other features, gaps, spirals and meridional flows that may be linked to the presence of planets \citep[see e.g. review by][]{pinte2022}.
These observations present an excellent testbed for theoretical models of planet-disc interactions. Interpreting them, however, requires a complex understanding of all the different processes at play.

During the last decades, observational campaigns spanning multiple star-forming regions have shown that most protoplanetary discs have a lifetime of only a few million years \citep{haisch2001, mamajek2009, fedele2010, ribas2014}.
Moreover, the observed disc fractions at different wavelengths suggest that most discs disperse from the inside-out \citep[see also][]{koepferl2013, ercolano2015a}.
While many physical processes have been proposed to drive the dispersal of discs, it is still mostly unconstrained which process is dominant at a given location and time.
It is clear that accretion onto the star plays a major role in the disc's evolution, but the mechanism that is responsible for the necessary transport of angular momentum is still uncertain and could be either turbulence driven by various types of instabilities or magnetic winds or a combination of both \citep[e.g.][for a recent review]{lesur2022}. Very likely, the dominant process varies with the location in the disc and its evolutionary stage.
In recent years, magnetically driven disc winds have been shown to be able to drive mass-loss and accretion at rates compatible with observations \citep[e.g.][]{gressel2015, bai2016a, bethune2017, wang2019, gressel2020, lesur2021}, adding to a growing body of evidence that they play an important role in disc evolution, especially at early stages, while later in the disc's life, thermal winds may become the dominant dispersal mechanism \citep[see also][]{ercolano2017, pascucci2020, pascucci2022}.
Models of thermal winds, in particular those driven by internal photoevaporation, have been very successful in explaining the lifetimes and fast inside-out dispersal of protoplanetary discs \citep[e.g.][]{alexander2006, gorti2009, owen2010, picogna2019}, as well as numerous other observables, including some of the observed spectra of wind-tracing emission lines \citep{ercolano2010, ercolano2016, weber2020, rab2022}, the spatially resolved [O\textsc{i}]~6300~\AA{} spectral line in TW Hya \citep{rab2023} or stars with low accretion rates \citep{ercolano2023}. We refer to \citet{ercolano2017, ercolano2022, pascucci2022} for recent reviews on the topic.

Disc winds not only play an essential role in the disc's dispersal, but they can also strongly affect the evolution of planets and vice versa.
Multiple theoretical works have shown that photoevaporative winds can strongly affect the orbital migration and final semi-major axis distribution of planets \citep{alexander2009, alexander2012, ercolano2015b, jennings2018, monsch2021a}.
Additionally, in an observational study, \citet{monsch2019} have constructed a catalogue of the X-ray luminosities and stellar and planetary properties of stars hosting giant planets and found a void in the distribution of giant planets in the plane of semi-major axis versus X-ray luminosity, which could point towards the parking of giant planets close to the location of a gap that is opened by X-ray photoevaporation \citep{monsch2021b}.

Moreover, an accreting planet that is massive enough to carve a gap can significantly reduce the mass flux across the gap, effectively starving the inner disc of material \citep[e.g.][]{lubow2006}. As a result, the wind-driven dispersal of the disc interior to the gap can be significantly accelerated, leading to the start of the transition disc phase, where the inner disc is depleted, and the then directly illuminated outer disc is quickly photoevaporated from the inside out \citep{alexander2009, rosotti2013}.
This effect is sometimes referred to as planet-induced photoevaporation (PIPE).
In a companion work \citep{weber2022}, we showed that a massive, gap-opening planet strongly affects the kinematics of a photoevaporative wind, an effect that could potentially be observed in emission line diagnostics as periodic variations in the spectral line profile with peaks in the redshifted part of the spectrum that trace the asymmetries generated by the planet in the wind.

In recent years, multiple theoretical works have shown that magnetic winds, too, can strongly influence the gap-opening, planet migration and accretion onto planets \citep[e.g.][]{kimmig2020, lega2021, elbakyan2022, nelson2023}.
\citet{kimmig2020} found that a planet's migration behaviour can differ strongly depending on the radial density gradients in the disc, with rapid type-III-like outward migration being possible under certain circumstances.
\citet{lega2021} showed that migration can be fast or slow, depending on whether or not the torque from the planet is strong enough to block the accretion flow and in a similar study \citet{nelson2023} demonstrated that accretion rates onto the planet itself also depend on this criterion.
\citet{elbakyan2022} found that it is easier to open deep gaps in discs with magnetised winds than in a standard viscous disc.
\citet{aoyama2023} carried out global three-dimensional non-ideal magnetohydrodynamic (MHD) simulations of planet-disc interactions, including magnetic winds and found that the magnetic flux concentrates in the corotation region of the planet, leading to a strongly enhanced angular momentum extraction from the gap region and an outflow that primarily originates from the disc surface outward of the gap. Moreover, their gaps are wider and deeper than in the purely viscous or inviscid models, reducing the torque acting on the planet. 
With a different 3D global non-ideal MHD model, \citet{wafflard-fernandez2023} recently found that the planet deflects magnetic field lines at the disc surface, resulting in a more efficient wind at the outer edge of the gap and less efficient wind at the inner edge, which creates an asymmetric gap and consequently an asymmetric torque acting on the planet, leading to slower or potentially even outward migration.

This highlights the importance of understanding planet-disc interactions and disc winds not just as two separate physical processes but as an intertwined system with feedback processes that play a crucial role in the evolution of newly formed planets. In this work, we aim to study the interplay between planet-disc interactions and photoevaporative disc winds, expanding upon prior work in this field by conducting global, three-dimensional radiation-hydrodynamic simulations of a protoplanetary disc hosting a Jupiter-like planet and undergoing X-ray photoevaporation.

In Sect. \ref{sec:methods}, we describe the details of our numerical model and the assumptions made.
In Sect. \ref{sec:results}, we present our results, followed by a discussion in Sect. \ref{sec:discussion} and a summary in Sect. \ref{sec:summary}.

\section{Methods}\label{sec:methods}
\subsection{Model strategy and initial conditions}
To reduce the computational cost of our simulations, we divide our models into three stages.
In stage I, we start with a two-dimensional model of a purely viscously evolving protoplanetary disc in the midplane ($R-\Phi$ plane), to which we add a Jupiter-like planet that is held fixed on its circular orbit.
The model is then evolved until the substructures that are generated by the planet have reached a steady state in the gas density.
In stage II, we extend the model into three-dimensional space by adding a vertical dimension extending to about six disc scale heights to capture the planet-disc interactions fully. 
Again, the model is evolved until a quasi-steady state has been reached.
In stage III, we activate the accretion of gas onto the planet and extend the 3D grid in the vertical direction to about 29 disc scale heights to prepare for the inclusion of a disc wind. 
Moreover, we branch off the model. In the first branch, we switch on the temperature parametrisation by \citet{picogna2019} to model the temperature structure of the disc's upper layers in the presence of EUV and X-ray irradiation from the central star, which results in a photoevaporative wind.
In the second one, we keep evolving the disc purely viscously, which is achieved by turning off the temperature parametrisation by setting the X-ray luminosity $L_X$ to zero.
All other physical parameters of the model remain the same across all stages.
This allows us to directly compare the effect of a thermal wind on the disc and planet evolution.
From now on, we will refer to the purely viscous model (without wind) as \textsc{nowind} and to the model with the temperature parametrisation enabled as \textsc{pewind}.
An extensive list of parameters and their values can be found in table \ref{tab:model-params} and we list here only the most important ones, namely the stellar mass $M_* = 0.7~M_\odot$, the viscosity parameter $\alpha = 6.9\cdot10^{-4}$, the planet mass $M_P = M_J$ and the semi-major-axis $a = 5.2~au$.
The details for the individual stages are described in the following subsections.

\begin{table}
\centering
\caption{Parameters for the hydrodynamic models. Common parameters are shared between all models.}
\label{tab:model-params}
\resizebox{.48\textwidth}{!}{
    \begin{tabular}{llll}
        \toprule
        \textbf{Model}  & \textbf{Parameter}    & \textbf{Value}            & \textbf{Description} \\
        \midrule
        Common          & $M_*$                 & 0.7~M$_\odot$             & stellar mass \\
                        & $M_P$                 & 1~M$_\mathrm{J}$          & planet mass \\
                        & $a_P$                 & 5.2~au                    & semi-major axis \\
                        & $\alpha$              & $6.9\cdot10^{-4}$         & viscosity parameter \\
                        & $\gamma$              & 1.67                      & adiabatic index \\
                        & $\mu$                 & 2.35                      & mean molecular weight \\
                        & $\tau_\mathrm{damp}$  & 0.3                       & damping timescale \\
        \hline          
        Stage I         & $\Sigma_0$            & 30~g\,cm$^{-2}$           & initial surface density \\
                        & $p$                   & 1.0                       & surface density slope \\
                        & $h$                   & 0.05                      & aspect ratio \\
                        & $f$                   & 0.25                      & flaring index \\
                        & $s$                   & 0.6                       & smoothing length \\
                        & $t_\mathrm{taper}$    & 20                        & mass tapering time \\
                        & $l_\mathrm{damp}$     & 1.15                      & damping zone width \\                    
        \hline  
        Stage II \& III & $w_\mathrm{damp}$     & 0.4~au                    & damping zone width \\ 
                        & $r_{sm}$              & 0.5~R$_\mathrm{H}$        & smoothing length \\
                        & $\rho_\mathrm{floor}$ & $10^{-21}$~g\,cm$^{-3}$   & floor density \\
        \hline  
        Stage III       & $\tau_\mathrm{acc}$   & 0.5                       & accretion timescale \\
        \hline      
        \textsc{pewind} & L$_X$                 & $2 \cdot 10^{30}$ erg\,s$^{-1}$   & X-ray luminosity \\
        \bottomrule
    \end{tabular}
}
\end{table}

\subsection{2D setup}
To carry out our initial two-dimensional simulations of the planet-disc interactions, we use the hydrodynamics version of the \textsc{fargo3d} code \citep{benitez-llambay2016a} with the orbital advection algorithm of \citet{masset2000}.
We use a polar grid with 776 uniformly spaced zones in the azimuthal and 512 logarithmically spaced zones in the radial direction between 0.364~au and 26~au, centred on the star with $M_* = 0.7~M_\odot$. An overview of all grids used in the different stages is given in table \ref{tab:model-grids}.
In the initial setup, the surface density $\Sigma$ is set by the profile
\begin{equation}
\Sigma(R) = \Sigma_0(R) \left( \frac{R}{R_0}  \right)^{-p}.
\end{equation}
For the initial surface density, we choose $\Sigma_0 = 30~g~cm^{-2}$ at the base length unit $R_0 = 5.2~au$ and a slope $p = 1$.
Further, the setup assumes an isothermal, mildly flaring disc with flaring-index $f=0.25$ that sets the aspect ratio according to $h(R) = H/R = 0.05~(R/R_0)^f$.
We include viscosity using the well-known $\alpha$ parametrisation by \citet{shakura1973} with $\alpha = 6.9\cdot10^{-4}$.
In the azimuthal direction, periodic boundary conditions are applied.
In the radial direction, the azimuthal velocity and density are extrapolated using a Keplerian profile and the initial surface density profile, respectively, and an antisymmetric boundary condition is used for the radial velocity.
To prevent oscillations at the boundary, we apply the wave-dampening method by \citet{deval-borro2006} with the timescale $\tau_{\mathrm{damp}} = 0.3$ and a width parameter $l_\mathrm{damp}$ (DampingZone) of 1.15, roughly corresponding to 10 per cent of the radius at the boundary.

The planet is modelled by including an additional gravitational potential of the form 
\begin{equation}
    \Phi_P = \sqrt{\frac{GM_P}{R^2 + s^2}},
\end{equation}
with a smoothing length defined by $s = 0.6~h(R)~R~$ (ThicknessSmoothing).
Since we use a reference frame that is corotating with the planet, we also include the indirect term that arises from the star's acceleration by the planet.
We assume a circular orbit with a fixed semi-major axis $a_P = 5.2~au$ and fixed mass $M_P = M_J$. 
To prevent shocks from inserting a massive planet into an unperturbed disc, we slowly increase its mass to its final value during the first 20 orbits.
The model is run for 1000 orbits until a steady state is achieved.

\begin{table*}
\centering
\caption{Grids used in the hydrodynamic models. Radii are given in code units where the unit length is 5.2~au.}
\label{tab:model-grids}
\resizebox{.85\textwidth}{!}{
\begin{tabular}{l|lll|lll|lll}
    \toprule
    \multicolumn{1}{c}{} & \multicolumn{3}{c}{\textbf{Stage I}} & \multicolumn{3}{c}{\textbf{Stage II}} & \multicolumn{3}{c}{\textbf{Stage III}}  \\
    \cmidrule(rl){2-4} \cmidrule(rl){5-7} \cmidrule(rl){8-10}
    \textbf{Dim.} & Interval & N Points & Type & Interval & N Points & Type & Interval & N Points & Type \\
    \midrule 
        & & &                                   & [0.280, 0.846] & 60 & uniform         & [0.280, 0.846] & 60 & uniform \\
    r   & [0.07, 5] & 512 & logarithmic         & [0.846, 1.154] & 80 & uniform         & [0.846, 1.154] & 80 & uniform \\
        & & &                                   & [1.154, 2.500] & 60 & logarithmic     & [1.154, 2.500] & 60 & logarithmic \\
    \hline
    
                & & & & [1.274, 1.417] & 15 & uniform                & [0.600, 1.040] & 26 & uniform \\
    $\theta$    & & & & [1.417, $\frac{\pi}{2}$] & 40 & uniform      & [1.040, 1.416] & 29 & uniform \\
                & & & &                                          & & & [1.416, $\frac{\pi}{2}$] & 40 & uniform \\                                                  
    \hline
    
            & & &                                   & [0, 2.988] & 244 & uniform            & [0, 2.988] & 244 & uniform \\
    $\Phi$  & [-$\pi$, $\pi$] & 776 & uniform       & [2.988, 3.295] & 80 & uniform         & [2.988, 3.295] & 80 & uniform \\
            & & &                                   & [3.295, 2$\pi$] & 244 & uniform       & [3.295, 2$\pi$] & 244 & uniform \\
    \bottomrule
\end{tabular}
}
\end{table*}

\subsection{3D setup}\label{sec:met:3dsetup}
We use the results of our initial 2D simulation to inform the setup of our 3D model by mapping the density and azimuthal velocity information from the 2D model into a 3D spherical grid.
For the density, we assume hydrostatic equilibrium, while the azimuthal velocity is set according to 
\begin{equation}
v_\phi = v_K \left[ (-p-q) \left( \frac{H}{R} \right)^2 + (1-q) - \frac{-qR}{\sqrt{R^2 + Z^2}} \right]^{1/2},
\end{equation}
which takes into account the force balance in the radial and vertical directions \citep[e.g.][]{nelson2013}. Here, $R$ and $Z$ are the cylindrical radius and height above the midplane, respectively, and $q = 0.5$ defines the slope of the temperature profile $T(R) = T_0 (R/R_0)^{-q}$ and can also be expressed as $q = 1 - 2f$. 
The radial and polar velocities are set to zero.
Although this does not take into account local deviations due to the planet and its substructures, it is sufficiently accurate for reducing the time required for the substructures in the 3D model to reach a quasi-steady state in the gas-density, compared to a start from a smooth disc without any substructure.

We perform all three-dimensional simulations using a modified version of the \textsc{pluto} code \citep{mignone2007}.
The code switch is merely for convenience, as our photoevaporation model is already implemented in \textsc{pluto}, whereas the 2D model was quicker to set up in \textsc{fargo3d}.
The 3D models are run in a spherical grid, spanning in the radial direction from $r = 1.456~au$ to $r = 13~au$ and in stage II (stage III) from $\theta = 1.274~(0.6)$, corresponding to $\approx 6~(29)$ scale heights at the location of the planet, to $\theta = \frac{\pi}{2}$ (the disc midplane).
We use a nested static grid to achieve a resolution of $\approx 0.05 R_{H}$ inside two Hill radii $R_{H}$ from the planet.
The details of the grids are listed in table \ref{tab:model-grids}.
To further reduce the computational cost, we assume symmetry with respect to the midplane, which allows us to model only half of the disc. 
To do this, we choose our boundary conditions in the polar direction to be reflective at the midplane. 
On the other side, we use a reflective boundary in stage II and an open boundary in stage III to allow the wind to escape the domain.
We note that the open boundary may cause reflections when the outflowing gas is subsonic, which is indeed the case in most parts of the simulation domain.
However, we verified that the structure of the wind does not differ in a significant way from that in the previous generation of models presented in \citet{weber2022}.
In that work, the focus lay on the inner regions of the wind, thus a much larger domain with the upper polar boundary very close to the polar axis was used.
In the radial directions, we again extrapolate the azimuthal velocity component using a Keplerian profile and choose an open boundary condition for the other quantities. 
At the azimuthal boundary, periodic conditions are applied.
As in the 2D model, we apply the wave-dampening method by \citet{deval-borro2006} with a width $w_\mathrm{damp}$ = 0.4~au and timescale $\tau_{\mathrm{damp}} = 0.3$.
Viscosity is implemented using \textsc{pluto}'s default implementation with zero bulk viscosity. This means that the stress tensor is calculated according to:
\begin{equation*}
    \vec{\tau} = \rho \nu \left[ \nabla \vec{v} + (\nabla \vec{v}^\mathrm{T}) - \frac{2}{3} (\nabla \cdot \vec{v}) \vec{I} \right],
\end{equation*}
where $\vec{I}$ is the identity matrix and $\nu = \alpha c^2_s \Omega^{-1}_K$ is the kinematic viscosity with $c_s$ being the sound speed and $\Omega_K$ the Keplerian angular velocity.

As in the 2D model, the planet is implemented as an additional gravitational potential, where we adopt the potential from \citet{klahr2006}, which, in 3D, is more accurate than the classical smoothed potential that we used for the 2D model:
\begin{align}
    \Phi_P =   
    \begin{cases}
        -\frac{G~M_P}{d} \left[ \left( \frac{d}{r_{sm}}\right)^4 - 2 \left( \frac{d}{r_{sm}}\right)^3 + 2 \frac{d}{r_{sm}} \right],& \text{if } d\leq r_{sm}\\
        -\frac{G~M_P}{d},              & \text{if } d > r_{sm}
    \end{cases}.
\end{align}
Here, $d$ is the distance from the planet and $r_{sm}$ the smoothing length, which we set to 0.5~$R_H$, such that the potential will be smoothed if the distance to the planet is less than 0.5 Hill radii.
Again, we use a rotating reference frame and add the indirect term $\Phi_{Ind,P} = -GM_P/a_P$ to the potential to account for the acceleration of the star due to the planet.
The planet is fixed on its circular orbit and does not accrete in stage II.

We run the model in stage II for 444 orbits, after which we extend the grid in the vertical direction to that of stage III. This will allow us to include a significant portion of the disc wind that will be launched in stage III.
After another 40 orbits, we activate the accretion onto the planet, which is detailed in Sect. \ref{sec:met:planet-accretion}, and branch off the model into the two branches with photoevaporation turned on or off.
Photoevaporation is implemented via the temperature parametrisation by \citet{picogna2019}. The details for it are explained in the following subsection.
The two branches are then run for another 200 orbits.
To trace gas flows in the disc and wind, we add passive scalar tracers after the first 100 orbits of stage III; the details for those are described in Sect. \ref{sec:met:tracers}.

\subsubsection{Temperature structure and photoevaporation model} \label{sec:met:photoevap-model}
When modeling thermal winds, it is important to accurately determine the temperatures in the wind and at the disc-wind interface.
To this end, we implement the temperature parametrisation by \citet{picogna2019} for a photoevaporative wind driven by EUV + X-ray irradiation from the central star.
The parametrisation is derived for a 0.7~$M_\odot$ star that irradiates the disc with the synthetic EUV + X-ray spectrum presented by \citet{ercolano2008a, ercolano2009} with an X-ray luminosity of $L_X = 2\cdot10^{30}$~erg/s.
The disc is assumed to have solar abundances \citep{asplund2005}, depleted according to \citet{savage1996} to account for the gas locked in dust grains.
With this setup \citet{picogna2019} carried out detailed radiative transfer calculations with the gas photoionisation and dust radiative transfer code \textsc{mocassin} \citep{Ercolano2003, Ercolano2005, Ercolano2008b}, which includes all relevant microphysical processes to self-consistently solve the ionisation and thermal balance of the gas and dust.
From the result, they then created a temperature parameterisation that depends only on parameters that are easily accessible during the hydrodynamic simulation: the column number density $N_H$ along the line-of-sight towards the star and the local ionisation parameter $\xi = L_X / nr^2$, where $n$ is the local number density and $r$ the distance to the star.
It is applied where the column density is less than $2.5\cdot10^{22}$~cm$^{-2}$, which is the maximum penetration depth of X-rays \citet{ercolano2009}, and it covers the wind and disc-wind interface.
At higher column densities, it is assumed that the irradiation from the star is screened and the gas temperatures are coupled to the dust temperatures, which are adopted from \citet{dalessio2001}. 
For more details, we refer to \citet{picogna2019}.

The advantage of this parameterisation is that it enables efficient temperature updates during every hydrodynamic timestep while maintaining high accuracy to the full radiative transfer calculation.
The disadvantage is that it neglects any additional heating by the planet or its substructures, which could have implications on the gas dynamics, especially close to the planet, e.g. for material that is accreting onto the planet. 
However, since our Jupiter-mass planet is well in the regime of runaway gas accretion, where the accretion rate is only limited by the amount of gas the disc can supply, the exact temperature structure and gas dynamics close to the planet are less important.
We therefore approximate the temperature in the vicinity of the planet using a simple 1D radial power-law profile with an exponential cut-off that is similar to the profile found by \citet{szulagyi2017} for a Jupiter-mass planet with a temperature of 4000~K:
\begin{equation}
    \resizebox{.45\textwidth}{!}{
    $T(d) = max \left\{ ~T_\mathrm{dust},~ 230~K \cdot \left( \frac{d}{R_H} \right)^{-0.62} \left( 1 - \frac{1}{exp(2-\frac{2d}{R_H}) + 1} \right) \right\},$
    }
\end{equation}
where $d$ is the distance to the planet.
Another shortcoming of this approach is that it neglects heating in shaded regions from a diffuse EUV or X-ray field that could change the temperature structure in the planetary gap.
To verify that this effect is not strong enough to affect our results significantly, we post-processed our model a posteriori with \textsc{mocassin} to find the temperature structure in the shaded gap region when the diffuse field is accounted for.
We present the details of this test in appendix \ref{sec:appendix-Tgap}.

\subsubsection{Gas accretion by the planet} \label{sec:met:planet-accretion}
To model the accretion of gas onto the planet, we use a simple method that is similar to the approaches described by \citet{dangelo2008} and \citet{durmann2017}.
At each hydrodynamic time step $\Delta t$, we remove an amount of gas within $0.5~R_H$ of the planet, which is given by $\frac{\Delta t}{\tau_\mathrm{acc}} \int \rho dV$, where $\tau_\mathrm{acc}$ can be understood as an accretion timescale.
We choose $\tau_\mathrm{acc} = 0.5~\Omega_P^{-1}$ inside $0.5~R_H$ and $\tau_\mathrm{acc} = 0.25~\Omega_P^{-1}$ inside $0.25~R_H$, which means that the accretion is twice as efficient in the inner quarter of the Hill sphere.
For simplicity, we only remove the accreted mass from the disc without adding it to the planet, which is justified considering that our simulation will be evolved only for short timescales and the total accreted mass will be small compared to the planet's mass.

This accretion method, while simplistic, is sufficient to estimate accretion rates in the runaway gas accretion scenario that we want to model here. It is not our focus to study the detailed dynamics of accretion flows but to estimate the influence that a wind can have on the accretion rate due to the redistribution of disc material.
More accurate modelling would require higher resolution and more sophisticated treatment of the heating and cooling processes in the vicinity of the planet. However, this would be too computationally demanding for our global model.

\subsubsection{Gas tracers} \label{sec:met:tracers}
To trace and quantify the radial redistribution of gas in the models, we add multiple passive scalars $Q_k$ as tracer variables. They obey the advection equation
\begin{equation}
    \frac{\delta \rho Q_k}{\delta t} + \nabla \cdot (\rho Q_k \vec{v}) = 0,
\end{equation}
where $k$ is the index of the tracer, $\rho$ the gas density and $\vec{v}$ the velocity vector.
The tracers are placed between 2 and 10~au in radial bins of width 0.5~au that span the entire vertical and annular domain. This is done by setting $Q_k = 1$ within the radial bin and $Q_k = 0$ elsewhere so that each tracer initially traces 100 per cent of the mass in its bin. We add the tracers only after the first 100 orbits of stage III to ensure that the photoevaporative wind is fully established at their introduction. As a result, the initial amount of mass that they trace differs between the models. To address this problem, we present in our results also a version where we perform our analysis after weighing the \textsc{pewind} model with the fraction of the initially traced mass,
\begin{equation}
    w_k = \frac{M_{k,\textsc{nowind}}}{M_{k,\textsc{pewind}}}.
\end{equation}

\section{Results}\label{sec:results}
In this section, we present the results of our simulations. 
Unless stated otherwise, or unless we present a temporal evolution, we show averaged quantities calculated from 100 snapshots during the last ten orbits of the simulation (orbit 190 to 200 for the \textsc{nowind} and \textsc{pewind} models). When mentioning a radius or radial components, we refer to the cylindrical radius (capital $R$) unless we explicitly refer to the spherical radius (lowercase $r$).

\subsection{Surface density structure and evolution}\label{sec:res:sigma}
\begin{figure}
    \resizebox{\hsize}{!}{\includegraphics{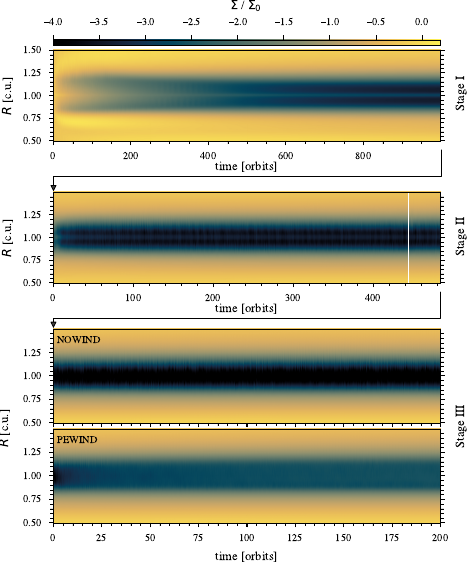}}
    \caption{Surface density evolution during the different model stages, normalized by the initial surface density profile. From top to bottom: stage I (2D model), stage II (start of 3D model), and stage III with \textsc{nowind} and \textsc{pewind} model, respectively. The white vertical line in the second panel indicates the time at which the model was vertically extended to the final 3D grid (see Sect. \ref{sec:met:3dsetup} for details). A black arrow between two panels means that the following stage was started from the end of the previous stage. We note that the scale of the time axis varies between the panels.}  
    \label{fig:sigma-evol}
\end{figure}
Fig. \ref{fig:sigma-evol} shows the one-dimensional gas surface density evolution for all model stages. 
It can be seen that in stage one (the initial 2D model), the planet opens up a deep gap over the course of a few hundred orbits. In the middle of the gap, where the planet is located, remains a bump in the surface density from gas accumulating in the circumplanetary region. 
The surface density reaches a steady state after 600 -- 800 orbits. 

After 1000 orbits, we extend the model into three dimensions and start stage II. During the first $\approx40$ orbits, the surface density adjusts to the new 3D structure before it, too, reaches a steady state. The width and depth of the gap in the 2D and 3D models are similar.  
After 444 orbits in stage II (marked by the white vertical line in the second panel of Fig. \ref{fig:sigma-evol}), we again extend the domain in the vertical direction to allow the disc wind to be included in the next stage. 
Since the newly created space is virtually empty with the gas density set to the floor density, the model does not need to adjust, which we verify by running it for another 40 orbits.

\begin{figure}
    \resizebox{\hsize}{!}{\includegraphics{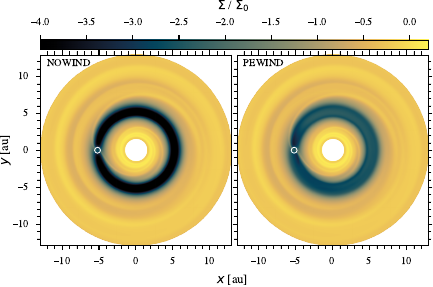}}
    \caption{Surface density maps of the \textsc{nowind} (top panel) and \textsc{pewind} (bottom panel) models, normalized by the initial surface density profile. The white dashed circle indicates the Hill sphere of the planet.}  
    \label{fig:sigma2d}
\end{figure}
In stage III, we activate the gas accretion onto the planet and start another model branch, where the temperature parametrisation described in Sect. \ref{sec:met:photoevap-model} is also active.
We refer to the models with and without photoevaporation by the identifiers \textsc{pewind} and \textsc{nowind}, respectively. 
In both models, the surface density bump at the planet's location very quickly disappears as the planet undergoes runaway gas accretion.
While there is no change in the width of the gap in either model, in the \textsc{pewind} model, the depth of the gap reduces by more than an order of magnitude within less than 30 orbits after the start of stage III.

We present the two-dimensional structure of the stage III models in Fig. \ref{fig:sigma2d}, where it is visible that the reduction of the depth of the gap in the \textsc{pewind} model is relatively uniform, almost around the entire azimuth. 
Only close to the planet, where the gas is following horseshoe orbits, the surface density is lower than in the rest of the gap.
Outside the gap, the models remain comparable in their overall disc structure, including the width of the gap and the geometry of the spirals.

\begin{figure}
    \resizebox{\hsize}{!}{\includegraphics{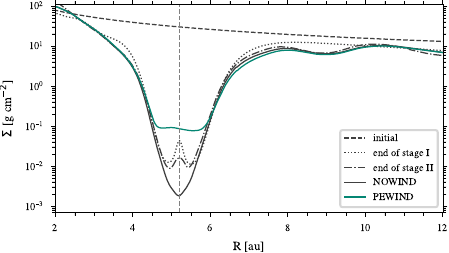}}
    \caption{Azimuthally averaged 1D surface density profiles. The grey dashed line indicates the location of the planet.}
    \label{fig:sigma1d}
\end{figure}
In Fig. \ref{fig:sigma1d}, we present the initial 1d gas surface density profile and the azimuthally averaged surface density profiles at the end of every stage in detail. Comparing the profile of the \textsc{pewind} model with that at the end of stage III confirms that the gap has become shallower by a factor of $\approx 10$ and even more when we compare it to the \textsc{nowind} model, where the gap has been depleted further by the accreting planet. The width of the gap is not affected. Outside the gap, the surface density stays similar between the 3D models, but the \textsc{pewind} model exhibits a slightly reduced surface density between approximately 6.5 and 8~au.

\begin{figure}
    \resizebox{\hsize}{!}{\includegraphics{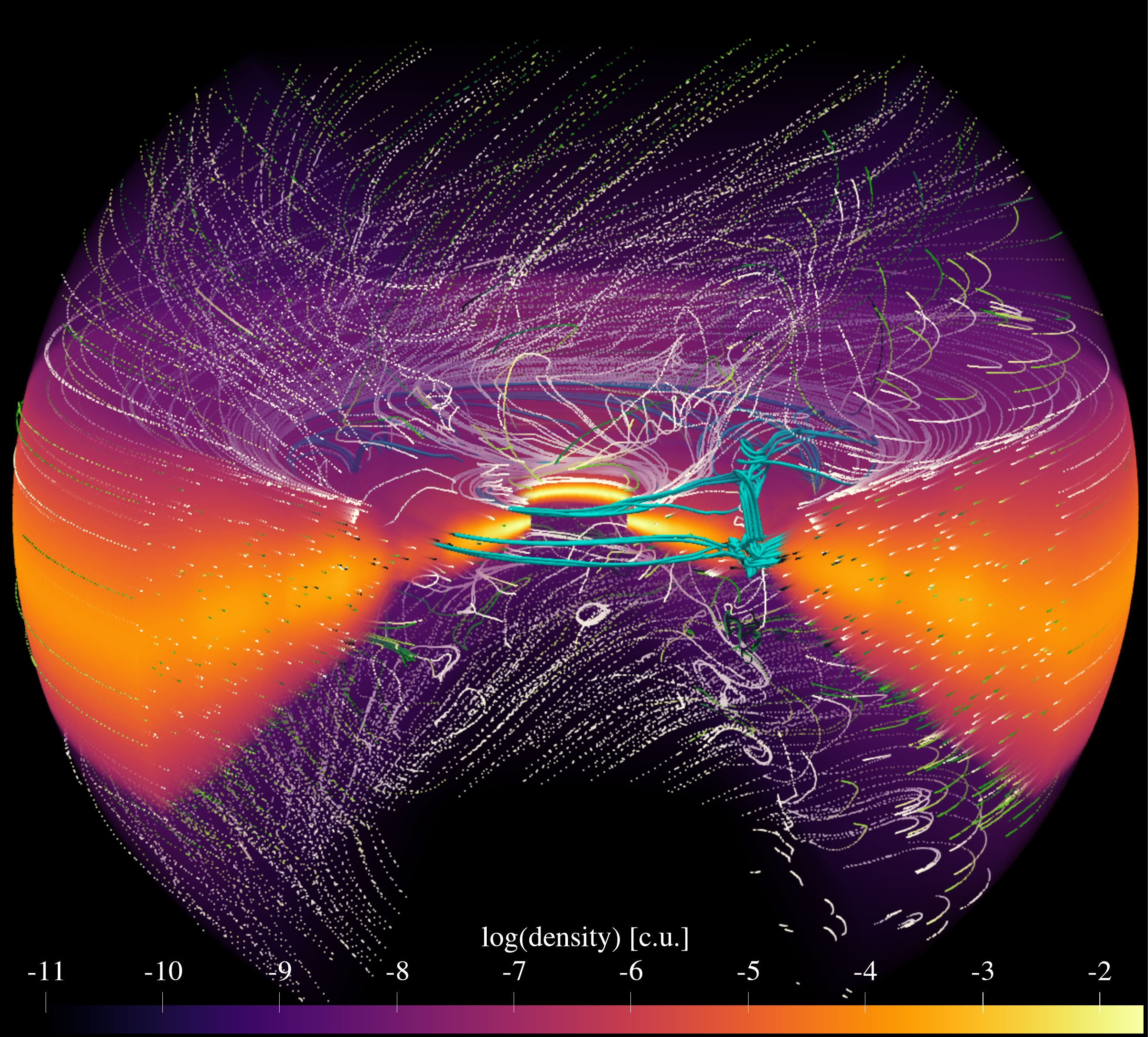}}
    \caption{3D visualization of the density in the \textsc{pewind} model created from a snapshot a t = 100 orbits. The dotted lines are gas streamlines. Their colour represents the integration time, which is useful for understanding the direction gas follows along the streamline (from white to green). The cyan tubes are gas streamlines traced back from within the planet's Hill sphere and indicate example paths that the gas can take towards the planet.}
    \label{fig:3dvis}
\end{figure}

Fig. \ref{fig:3dvis} shows a three-dimensional representation of the density in the \textsc{pewind} model, overlain by gas streamlines in the disc and wind (white), as well as streamlines that show example paths along which gas can be transported into the planet's Hill sphere (green).

\subsection{Disc structure} \label{sec:res:structure}
\begin{figure*}
    \centering
    \includegraphics[width=17cm]{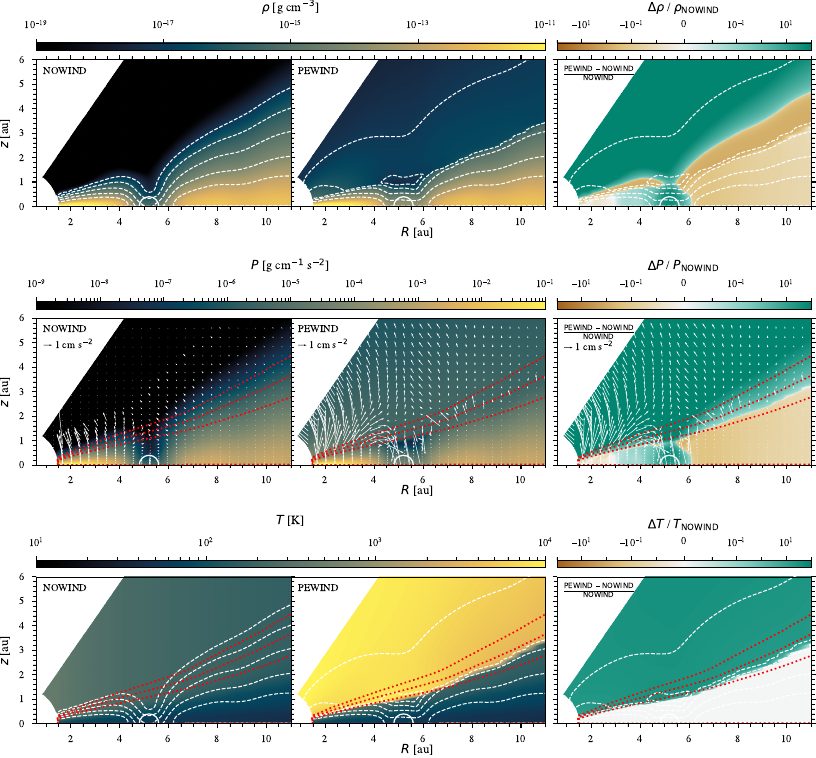}
    \caption{From top to bottom: Azimuthally averaged gas density, gas pressure and gas temperature. From left to right: \textsc{nowind} model, \textsc{pewind} model, and their relative difference. White dashed lines are density contours for (from top to bottom) $10^{-17}$ to $10^{-13}$~g\,cm$^{-3}$ in increments of one order of magnitude. Red dotted lines are column density contours for (from top to bottom) $2.5\cdot10^{20}$, $2.5\cdot10^{21}$, and $2.5\cdot10^{22}$~cm$^{-2}$. The arrows in the second row represent the pressure gradient force. In the right column, only the contours of the \textsc{pewind} model are shown. The white circles indicate the extent of the planet's Hill sphere.}  
    \label{fig:vertical-structure}
\end{figure*}
In previous work \citep{weber2022}, we have found that the planetary substructures, especially a gap in the disc gas, can profoundly affect the structure and kinematics of a photoevaporative wind. In the presence of a gap, the pressure gradient in the wind-launching region and even beyond the gap is significantly altered, and since a photoevaporative wind is a thermal wind driven by pressure gradient forces, the wind itself is strongly affected, too, which we can also observe in our \textsc{pewind} model. In this work, we are mainly interested in the implications of this for the disc and the planet. We will only briefly touch upon the impact on the wind and direct readers to our previous work for more details.

In Fig. \ref{fig:vertical-structure}, we show the azimuthally averaged density, pressure and gas temperature structure of the stage III models.
The panels in the top row show the density structure, overlain by density contours that indicate the gap profile.
The crucial difference between the models is the heating by EUV + X-ray irradiation from the central star that is applied in the \textsc{pewind} model via the temperature parametrisation. 
The panels in the third row show the resulting temperature structure, overplotted by column density contours up to $2.5\cdot10^{22}$~cm$^{-2}$, which is the maximum penetration depth of the X-rays.
At the heated surface layers of the disc, the \textsc{pewind} model experiences a sharp temperature rise (third row), accompanied by a sharp drop in the gas density, which is in contrast to the \textsc{nowind} model, where the density falls off exponentially in the vertical direction.
The transition from cold to hot gas can be understood as the transition from the bound disc to the wind-launching region, which happens at column number densities between $2.5\cdot10^{21}$ and $2.5\cdot10^{22}$~cm$^{-2}$.
A strong pressure gradient develops in this region, pointing away from the disc surface (second row), which can launch a dense wind.
However, at the location of the planetary gap, where the density and temperature of the bound disc are low, the pressure is also low.
As a result, the pressure gradient is pointing into the gap and no wind is launched from the location of the gap. We will discuss the structure of the wind in more detail in the next section.

As seen in the top-right panel, the disc in the \textsc{pewind} model has a slightly lower density (by a few per cent) radially outside the gap.
This can be explained by the mass that has been redistributed or lost due to the wind during the simulation.
We can confirm this with a quick estimation: The measured mass-loss rate due to the wind inside the domain is $\approx5\cdot10^{-9}~M_\odot~yr^{-1}$ (see Sect. \ref{sec:res:windmdot} for details). 
If the mass-loss rate were constant, we would expect to have lost $\approx 1.42\cdot10^{-5}M_\odot$ during our simulation.
The difference in total disc mass between the two models is $\approx1.37\cdot10^{-5}M_\odot$, which corresponds to $\approx2.5$ per cent of the mass at the start of stage III, consistent with the percent-level difference in the density in the outer disc.
Radially inside of the gap, as already noted in the previous section, the gas density is enhanced, and a region of increased density extends radially towards the star into the inner disc down to $R \approx 3$~au. 
A possible explanation could be that a fraction of the surplus gas inside the gap is viscously transported radially inward.
In that case, the planetary gap would be less efficient in blocking the accretion flows towards the star in a model that includes a photoevaporative wind.
We will study the details of this gas redistribution and its implications in the following sections.

\subsection{Wind structure and gas redistribution}\label{sec:res:redistribution}
\begin{figure*}
    \centering
    \includegraphics[width=17cm]{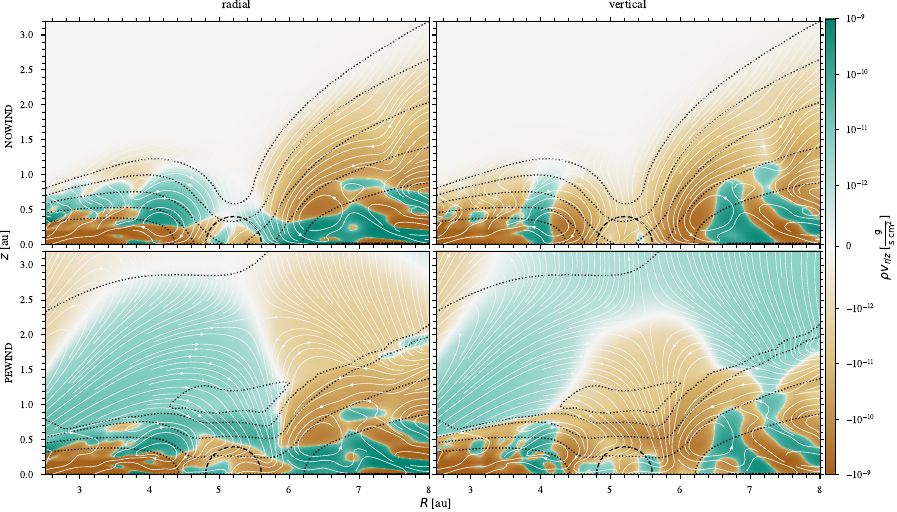}
    \caption{Azimuthally averaged radial and vertical mass flux. The dotted black lines are density contours for (from top to bottom) $10^{-17}$ to $10^{-13}$~g\,cm$^{-3}$ in increments of one order of magnitude. The dashed black circles show the Hill sphere of the planet. The white solid lines are streamlines of the gas.}
    \label{fig:rho-vrz}
\end{figure*} 
In Fig. \ref{fig:rho-vrz}, we present the radial and vertical components of the azimuthally averaged mass flux overlain by the gas streamlines.
The kinematic structure of the bound disc is very similar between the models, in agreement with our findings in the previous section that the wind has no significant impact on the bound disc outside of the gap, aside from slight differences in the density.
Particularly noteworthy are the meridional circulations that can be seen at the edges of the gap in both models.
The only exception is inside the gap above the planet's Hill sphere, where the \textsc{pewind} model has a lot of additional momentum directed radially outward and downward towards the midplane. This momentum is delivered to the gap by the wind.
The streamlines demonstrate that the wind is launched from the disc surface everywhere except at the location of the planetary gap.
This leads to a zone of low density and pressure in the wind above the gap, causing the pressure gradient in the surrounding wind to point towards that zone.
As a result, material that is launched into the wind from the inner disc is driven towards this zone, where it then falls back into the gap, and the same is true for material launched close to the outer edge of the gap.
The two flows then collide above the gap's outer edge, followed by the infall of wind material into the gap.
The collision of the opposing flows prevents the direct delivery of gas through the wind across the gap from the inner to the outer disc and vice versa.
Instead, mass transport across the gap happens only after the gas is delivered to the gap and subsequently pushed towards the edges by the planet's torque.

It is important to notice that this effect is entirely dominated by the presence of the gap and its implications on the pressure field and not due to the gravitational potential of the planet, which is negligible at the height of the disc surface.
This means that the described behaviour is similar around the entire azimuth and not limited to locations close to the planet.

Above the gap, at heights $z \approx 2$~au, is a turning point where the wind turns vertically upwards again, which is further reassuring that the wind structure in this region is not significantly affected by reflections at the open polar boundary.

\begin{figure}
    \resizebox{\hsize}{!}{\includegraphics{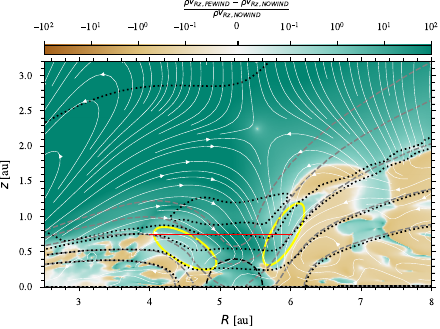}}
    \caption{Relative difference of the azimuthally averaged mass flux magnitude in the meridional plane. The dotted black lines are density contours of the \textsc{pewind} model for (from top to bottom) $10^{-17}$ to $10^{-13}$~g cm$^{-3}$ in increments of one order of magnitude. The dashed grey lines are the same contours for the \textsc{nowind} model. The red line defines the ring surface through which we calculate the vertical mass flux into the gap. The solid yellow ellipses highlight the gap edges with increased mass-flux in the \textsc{pewind} model. The dashed black circle shows the Hill sphere of the planet. The white solid lines are streamlines of the gas in the \textsc{pewind} model.}
    \label{fig:delta_rho-v}
\end{figure}
In addition to the gas that is delivered to the gap through the wind, the \textsc{pewind} model also has an increased radial and vertical mass flux directly at the gap edges.
This can be seen better in the relative difference of the mass flux magnitude within the meridional plane, which we show in Fig. \ref{fig:delta_rho-v}, where we marked these regions with yellow ellipses.
The increased flux in these regions is likely due to a combination of additional gas delivered by the wind and increased acceleration towards the gap by thermal pressure from the heated disc surface.
Moreover, as discussed in the previous section, the gas density is increased in the \textsc{pewind} model radially interior to the gap.
Consequently, the inner meridional flow carries more mass than in the \textsc{nowind} model, which is delivered into that region.
At the outer gap, the opposite effect would be expected, but the decrease in mass is outweighed by the mass delivered through the wind and the enhanced acceleration. 

These delivery paths combined result in a significant flux directly above the planet's corotation region, that flows towards the outer edge of the gap and vertically towards the midplane.
To quantify the amount of extra mass that is delivered through these pathways, we calculate the mass-flux through a ring surface defined by the red line in Fig. \ref{fig:delta_rho-v}. This surface spans across the gap, starting and ending where the $10^{-15}$~g\,cm$^{-3}$ density contours of both models meet, which roughly coincides with the borders of the region of enhanced mass-flux in the \textsc{pewind} model. 
For the \textsc{pewind} model, we find a rate of $\dot{m}_\mathrm{gap,v} \approx 10.7~M_J\,Myr^{-1}$ ($\approx1.0\cdot10^{-8}~M_\odot\,yr^{-1}$) flowing through this surface into the gap, compared to $\approx2.1M_J\,Myr^{-1}$ ($\approx2.0\cdot10^{-9}~M_\odot\,yr^{-1}$) in the \textsc{nowind} model.
We list the most important quantities measured in our models in table \ref{tab:results-quantified}.
We note that this measurement does not give an accurate picture of the total mass that flows into and out of the gap in both models, as it does not account for flows below this surface.
Instead, it serves to estimate the amount of gas delivered from the vertical direction.

\subsection{Accretion onto the planet}\label{res:sec:mpacc}
\begin{figure}
    \resizebox{\hsize}{!}{\includegraphics{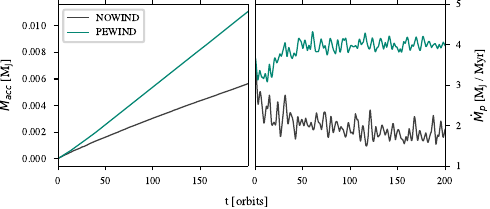}}
    \caption{Amount of mass accreted by the planet (left) and mass accretion rates onto the planet (right).}
    \label{fig:mpacc_mpdot}
\end{figure}
As discussed in the previous sections, the depth of the planetary gap is reduced in the \textsc{pewind} model by more than an order of magnitude. Therefore, the amount of mass available for accretion onto the planet is significantly increased.
In Fig. \ref{fig:mpacc_mpdot}, we present the temporal evolution of the accreted mass and the accretion rates onto the planet in both models, measured as the amount of mass removed from inside the Hill sphere, as described in Sect. \ref{sec:met:planet-accretion}.
After 200 orbits, the planets have accreted $5.6\cdot10^{-3}~M_J$ and $11.1\cdot10^{-3}~M_J$ in the \textsc{nowind} and \textsc{pewind} models, respectively.
For the mean accretion rate during the last 50 orbits, we measure $M_P \approx 1.8~M_J Myr^{-1}$ and $4.0~M_J Myr^{-1}$, showing that the accretion rate in the \textsc{pewind} model is more than twice the rate in the \textsc{nowind} model.
While significant, this increase in the accretion rate is much lower than the order of magnitude increase of the surface density in the gap.
However, as discussed in Sect. \ref{sec:res:sigma}, the density in the gap is not constant along the entire azimuth.
It can be seen in Fig. \ref{fig:sigma2d} that very close to the planet, the density is lower than far away from the planet.
This is because the gas within the coorbital region follows horseshoe orbits and does not necessarily reach the planet's Hill sphere.
In fact, the majority of the gas that reaches the Hill sphere usually does so from the vertical direction \citep{dangelo2003}, but the specific dynamics close to the planet and the structure of the circumplanetary disc depend strongly on the thermodynamics in the circumplanetary region \citep[e.g.][]{szulagyi2016b}, which we model only in the very basic manner described in Sect. \ref{sec:met:photoevap-model}.
Our measured accretion rates can, therefore, only provide a rough approximation.
It is nevertheless clear that photoevaporation increases the amount of mass available for accretion.

In the right panel in Fig. \ref{fig:mpacc_mpdot}, it can be seen that the planets in both models accrete with the same rate at the start of stage III, but while in the \textsc{nowind} model, the rate quickly drops after the initially available circumplanetary material is accreted, in the \textsc{pewind} model, photoevaporation can sustain the initial accretion rate already after only a few orbits before the accretion rate increases to its final value which is reached after $\approx50$ orbits.  

\begin{figure}
    \centering
    \resizebox{\hsize}{!}{\includegraphics{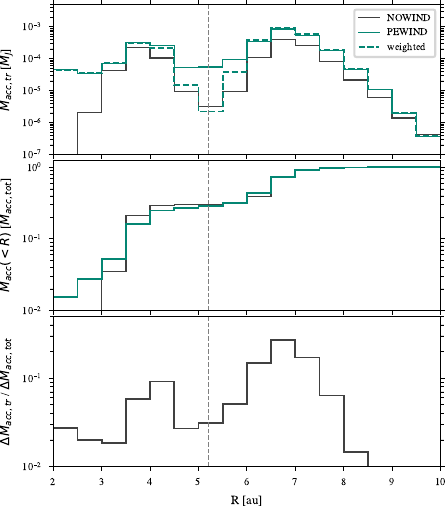}}
    \caption{Top panel: Contribution of the radial bins to the mass accreted by the planet, traced with passive scalars introduced at t=100 and evaluated at t=150 orbits. The dashed line shows the profile of the \textsc{pewind} model, weighted to account for the discrepancy in the initial amount of mass traced in each bin. 
    Middle panel: Cumulative radial profile.
    Bottom Panel: Contribution fraction of each radial bin to the additionally accreted mass in the \textsc{pewind} model.
    The grey vertical lines indicate the location of the planet.}
    \label{fig:mpacc-dr}
\end{figure}

To study which parts of the disc contribute most to the mass that the planet accretes, we added passive scalars to our simulations, allowing us to trace gas flows and determine the origin of the accreted gas.
Starting at orbit 100, we placed 16 passive scalars as tracers in radial bins of 0.5~au width between 2 and 10 au.
The bins span the entire verticle extent of the disc.
Fig. \ref{fig:mpacc-dr} shows the contribution of each radial bin to the accreted mass, the cumulative radial profile, and the fraction that each tracer contributes to the difference of total accreted mass between the models, all evaluated at orbit 150.
Since we place the tracers starting only at orbit 100, where there are already density differences between the models, the tracers do not trace the exact same amount of mass between the models.
However, this is only significant inside the gap, where the density difference is considerable. To show this, we also present the contribution of the radial bins in the \textsc{pewind} model after weighting them by the initially traced mass, which represents the most extreme case and can be considered a limiting case. This is shown as the dashed line in the top panel of Fig. \ref{fig:mpacc-dr}.
As is evident from the figure, the planet in the \textsc{pewind} model can accrete more material than in the \textsc{nowind} model from all radial bins up until 9.5 au, beyond which it accretes slightly less than in the \textsc{nowind} model.
This is consistent with the wind being less efficient in delivering material towards the gap from larger radii outside of the gap because, at large radii, the influence of the gap decreases up until the point where the wind is launched radially outward again, as would be expected if there was no gap. 

A particularly notable difference between the models is the contribution of the bins at $R < 3~au$:
It can be seen from the cumulative profile that in the \textsc{pewind} model, $\approx 3$ per cent of the total accreted mass originates inside three au, in contrast to the \textsc{nowind} model where there is no significant contribution from inside that radius.

\subsection{Planet Migration}\label{sec:res:migration}
\begin{figure*}
    \centering
    \includegraphics[width=17cm]{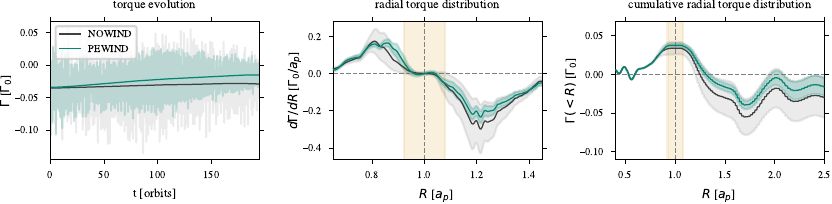}
    \caption{Torque evolution (left panel), radial torque distribution (middle panel) and cumulative radial torque (right panel). The temporal evolution was smoothed with a Gaussian kernel over ten orbits to make the plot more readable, and the shaded area indicates the spread of the actual samples. In the other panels, the shaded area indicates the standard deviation from the mean, calculated from 1000 samples between orbit 190 and 200. The yellow vertical band marks the extent of the planet's Hill sphere, and horizontal and vertical dashed lines indicate the zero point and the planet's location, respectively.}
    \label{fig:torque}
\end{figure*}
During our simulations, we also calculate the disc's gravitational torque that is acting on the planet by integrating over the entire disc, where we follow \citet{crida2008} and \citet{kley2009} to apply a tapering function
\begin{equation}
f_{\Gamma}(d) = \left[ exp\left( -\frac{d/R_H - b}{b/10} \right) \right],
\end{equation}
with $d$ being the distance from the planet and $b = 0.8$ a torque cut-off radius in units of $R_H$. This ensures that contributions from gas that is gravitationally bound to the planet are ignored, which is important because, in a more realistic model that does not neglect self-gravity, this gas should experience the same torque as the planet and therefore be considered a part of it.

During our analysis of the torques, we use the standard normalization \citep[][]{paardekooper2010}
\begin{equation}
\Gamma_0 = \Sigma_p \Omega^2_p a^4_p \left( \frac{M_P}{M_*} \right)^2 h^{-2}.
\end{equation}

The left panel of Fig. \ref{fig:torque} shows the temporal evolution of the torque.
Initially, the magnitude of the torques decreases over time in both models, but the decrease is faster in the \textsc{pewind} model. After $\approx 100$~orbits, the torque appears to reach a steady state in the \textsc{nowind} model, while it decreases further in the \textsc{pewind} model, albeit at a slightly reduced rate.
At the end of stage III, the torque in the \textsc{pewind} model is reduced almost by a factor of 2. If we assume that the torque remains constant in the \textsc{nowind} model, we can estimate the migration timescale
\begin{equation}
    \tau_M = \frac{a_p}{|\dot{a}|} = \frac{J_p}{2\Gamma} = \frac{M_p a_p^2 \Omega_p}{2\Gamma},
\end{equation}
where $J_p$ is the planet's angular momentum.
Evaluating the timescale for the \textsc{nowind} model, we find $\tau_{M,\textsc{nowind}} \approx 5.3\cdot10^{5}~yr$. 
For the \textsc{pewind} model, we find $\tau_{M,\textsc{pewind}} \approx 9.9\cdot10^{5}~yr$; however, the torque will likely continue to decrease in this model, slowing down the migration even more. 

As is best visible in the cumulative torque and the radial torque distribution, most of the difference between the models arises between 0.8 and 1.4$~a_p$, which corresponds to 4.16 and 7.28~au, but only a minor contribution comes from inside the corotation region, defined by the extent of the Hill sphere (marked yellow in the figure).
The difference can likely be explained by the enhanced density observed radially interior to the gap in the \textsc{pewind} model (see Sect. \ref{sec:res:structure}), leading to a higher positive torque around $R=0.9~a_p$.
The same argument holds the other way around: Radially outside the gap, the density is smaller, leading to a weaker negative torque around $R=1.2~a_p$.

\subsection{Mass transport across the gap}\label{sec:res:gapcross}
\begin{figure}
    \resizebox{\hsize}{!}{\includegraphics{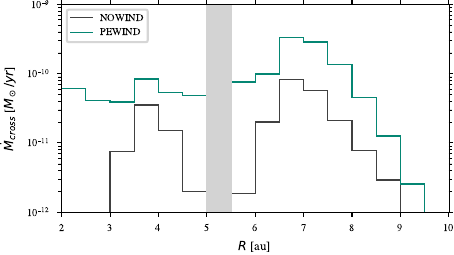}}
    \caption{Contribution of the radial bins to the rate of mass crossing the gap either outward (if to the right of the grey shaded area) or inward (if to the left of it).}
    \label{fig:mdot_gapcross}
\end{figure}
The transport of mass across a deep planetary gap can significantly be reduced by an accreting planet, depending on how efficient the planet is in accreting the gas that is supplied from disc \citep[e.g.][]{lubow2006}. 
It was proposed in the past that this could lead to the starvation of the inner disc, resulting in an accelerated dispersal of the inner disc, which could start the transition disc phase, where a photoevaporative wind quickly evaporates away the then fully exposed inner edge of the gas cavity \citep{alexander2009, rosotti2013}. 
This process is often called planet-induced photoevaporation (PIPE).
Using our three-dimensional models, we can investigate this scenario in full detail, taking into account not only the mass loss but also the redistribution of gas by the wind.

In Fig. \ref{fig:mdot_gapcross}, we present the mass rates crossing the gap either from the outside in or vice versa. 
The rates are calculated using our radially binned passive scalar tracers. Gas originating in a particular bin is considered to have crossed the gap if transported across the bin between 5.0 and 5.5 au (shaded in grey in the figure).
In the \textsc{nowind} model, the total rates of inward and outward transport are $\dot{m}_\mathrm{cross,in} \approx 2.0\cdot10^{-10}~M_\odot\,yr^{-1}$ and $\dot{m}_\mathrm{cross,out} \approx 6.1\cdot10^{-11}~M_\odot\,yr^{-1}$, respectively.
The \textsc{pewind} model has increased rates by factors of 5.4 and 5.1 with $9.9\cdot10^{-10}~M_\odot\,yr^{-1}$ and $3.3\cdot10^{-10}~M_\odot\,yr^{-1}$, respectively.
If we use the simple estimation for the steady state accretion rate onto the star in an unperturbed disc, $\dot{m}_{acc} = 3\pi \nu \Sigma$ with $\nu = \alpha c_s H$ and $H = \frac{c_s}{\Omega}$, we find $\dot{m}_{acc} \approx 6.6\cdot10^{-10} M_\odot\,yr^{-1}$. This rate is exceeded in the \textsc{pewind} model, while in the \textsc{nowind} model, only $\approx 30$ per cent of it can be sustained.
Moreover, we note that our reported values should be considered lower limits because our tracers are only placed between 2 and 10~au and mass further inside or outside is ignored in this calculation.
In particular, this could affect the outward transport rates of the \textsc{pewind} model, as the rates are still high for the innermost radial bin between 2 and 2.5 au.
Photoevaporation usually cannot launch a wind very close to the star ($R \lessapprox 2$~au) because the pressure gradients are not strong enough for the resulting force to overcome the strong gravitational force in the innermost regions of the disc.
However, it is nevertheless possible for gas from the inner disc to be pushed outward along the disc surface and later be picked up by the wind, a process that is facilitated by the presence of a gap that increases the pressure gradient in the radial direction \citep[see also][]{weber2022}.

\begin{figure}
    \resizebox{\hsize}{!}{\includegraphics{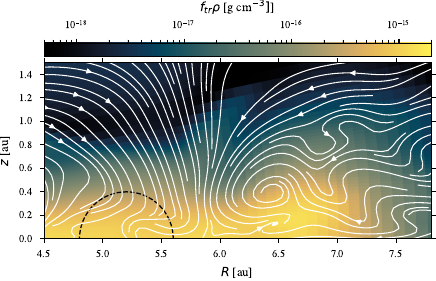}}
    \caption{Closeup view of the outer gap edge, showing the density of gas traced by the innermost tracer, i.e. of gas that originated between $R = 2$ and 2.5 au and was transported across the gap. Solid white lines are streamlines, and the black dashed circle indicates the extent of the planet's Hill sphere.}
    \label{fig:tracer-accumulation}
\end{figure}
In Sect. \ref{sec:res:redistribution}, we have seen that the wind cannot deliver material across the gap directly, because it collides with the opposing wind from the other side of the gap, leading to the infall of wind material into the gap.
However, the material delivered into the gap can indirectly cross the gap, when the torque from the planet pushes the material towards the edges of the gap, from where it can subsequently diffuse viscously.
To provide a more intuitive picture, we visualized the two flows from the outer and inner disc in appendix \ref{sec:appendix-tracers}, where we show three different snapshots of the innermost tracer in Fig. \ref{fig:stracer-0_evol} and of a tracer near the outer gap edge in Fig. \ref{fig:stracer-8_evol}.
Since the disc is viscously accreting onto the star, gas at the outer edge of the gap is not effectively transported farther outwards.
Instead, it accumulates at the edge of the gap.
In Fig. \ref{fig:tracer-accumulation}, we show the gas density traced by the tracer originating in our innermost radial bin between 2 and 2.5 au.
Between $R = 6.5$ and 7.5~au, at the location of the meridional flow, the traced gas appears to accumulate, with the tracer density quickly falling off at larger radii.

\subsection{Wind mass-loss rate and profile} \label{sec:res:windmdot}
\begin{figure}
    \resizebox{\hsize}{!}{\includegraphics{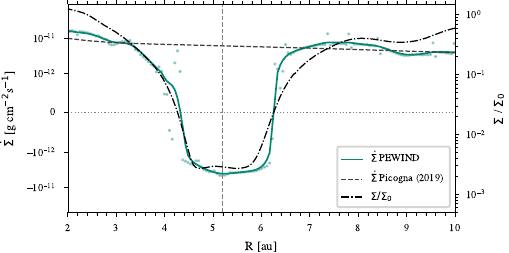}}
    \caption{Azimuthally averaged surface mass-loss profile of the \textsc{pewind} model. The coloured circles indicate the individual samples and the green solid line is obtained by smoothing the samples using a Gaussian kernel with $\sigma=3$. The black dashed line shows the surface mass-loss profile of the primordial disc by \citet{picogna2019}, which does not have any substructures; the vertical dashed line indicates the planet's location. The black dashed-dotted line shows the 1d surface density profile.}
    \label{fig:sigmadot-wind}
\end{figure}
In Fig. \ref{fig:sigmadot-wind}, we show the azimuthally averaged surface mass-loss profile in the \textsc{pewind} model.
The profile was calculated from the mass flux measured through the surface that is 0.4~au above the surface where the temperature gradient reaches its maximum (the transition from blue to yellow in Fig. \ref{fig:vertical-structure}).
Consistent with the streamlines discussed in the previous section, the profile shows an infall of mass between 4 and 6.4 au.
For comparison, we also show the surface mass-loss profile of the unperturbed primordial disc in \citet{picogna2019}. 
Although there are differences in the initial conditions between their setup and ours (primarily in disc mass and viscosity), and they use a different method to calculate the mass-loss profiles, the profiles should be robust against these differences since the wind mass-loss profile is mainly dependent on the X-ray spectrum and luminosity, the stellar mass and the disc aspect ratio, all of which we adopted from their model. 
Indeed, the two profiles reach similar mass-loss rates far away from the planetary gap. 
The total rate for mass that is lost through this surface (and not falling back) is $\dot{m}_\mathrm{wind} \approx 5.2\cdot10^{-9}~M_\odot yr^{-1}$. However, since the mass-loss close to the domain boundaries can be affected by the wave-dampening we apply, we limit our analysis to the radial range between 2 and 10~au, where the model experiences a wind mass-loss rate of $\approx4.1\cdot10^{-9}~M_\odot yr^{-1}$. Of that, $\approx1.1\cdot10^{-9}~M_\odot yr^{-1}$ is lost inside the planet's orbit. In the same radial range, the surface mass-loss profile of \citet{picogna2019} yields a rate of $\approx6.4\cdot10^{-9}~M_\odot yr^{-1}$ of which $\approx2.5\cdot10^{-9}~M_\odot yr^{-1}$ are lost inside 5.2 au.
The total rate for mass falling back into the gap is $\approx7.4\cdot10^{-10}~M_\odot yr^{-1}$. 
We note that this infall rate only includes gas lifted by at least 0.4 au from the surface and does not include material transported into the gap while staying below that threshold. It is therefore much lower than the rate we found in Sect. \ref{sec:res:redistribution}.

\begin{table*}
\centering
\caption{Overview of the most important quantities measured in the models.}
\label{tab:results-quantified}
\resizebox{.75\textwidth}{!}{
\begin{tabular}{lllll}
    \toprule
    \textbf{Quantity}   & \textbf{Description}  & \textbf{NOWIND}    & \textbf{PEWIND}   & \textbf{Unit} \\
    \midrule
    $\dot{M}_P$                     & accretion rate onto the planet            & $1.8\cdot10^{-6}$         & $4.0\cdot10^{-6}$         & $M_J yr^{-1}$ \\
    $\dot{m}_\mathrm{gap,v}$        & mass-flow vertically into the gap         & $0.2\cdot10^{-8}$         & $1.0\cdot10^{-8}$         & $M_\odot yr^{-1}$ \\
    $\dot{m}_\mathrm{cross,in}$     & mass-flow inwards across the gap          & $2.0\cdot10^{-10}$        & $9.9\cdot10^{-10}$        & $M_\odot yr^{-1}$ \\
    $\dot{m}_\mathrm{cross,out}$    & mass-flow outwards across the gap         & $0.6\cdot10^{-10}$        & $3.3\cdot10^{-10}$        & $M_\odot yr^{-1}$ \\
    $\dot{m}_\mathrm{wind}(2~au\leq R \leq 10~au)$  & wind mass-loss rate between 2 and 10 au &             & $4.1\cdot10^{-9}$         & $M_\odot yr^{-1}$ \\
    $\dot{m}_\mathrm{wind}(R < 5.2~au)$             & wind mass-loss rate inside 5.2 au       &             & $1.1\cdot10^{-9}$         & $M_\odot yr^{-1}$ \\
    $\tau_M$                        & migration timescale                       & $5.3\cdot10^5$            & $9.9\cdot10^5$            & $yr$ \\
    \bottomrule
\end{tabular}
}
\end{table*}

\section{Discussion}\label{sec:discussion}
\subsection{Inner disc lifetime and PIPE}
As shown in the previous sections, a significant fraction of the wind launched from the inner disc is recycled by falling back into the gap and then transported inwards again. 
In consequence, the wind mass-loss rate in the inner disc is reduced by a factor $\approx2.5$ to $\approx 1.1\cdot10^{-9}~M_\odot yr^{-1}$ (see Sect. \ref{sec:res:windmdot}). 
Additionally, due to the increase of mass that is loaded into the gap from outside of the planet's orbit and then pushed by the planet's torque towards the edges of the gap, an inward mass transport across the gap is sustained at a rate of $\approx 10^{-9}~M_\odot yr^{-1}$ (see Sect. \ref{sec:res:gapcross}).
This rate is about 50 per cent higher than the viscous accretion rate in the unperturbed disc despite the presence of a strongly accreting planet.
If we assume that all rates remain constant and that the star always accretes gas at the viscous rate, $\dot{m}_\mathrm{visc}$, we can make a simple estimate of the lifetime of the inner disc, $\tau_\mathrm{inner}$:
\begin{equation*}
    \tau_\mathrm{inner} = \frac{M_\mathrm{inner}}{\dot{m}_\mathrm{visc} + \dot{m}_\mathrm{wind}(R < 5.2~au) - \dot{m}_\mathrm{cross,in}}.
\end{equation*}
$M_\mathrm{inner}$ is the mass of the inner disc.
This is, of course, not an accurate estimate.
In reality, the rates will not remain constant over the entire lifespan.
For example, the accretion rate decreases with decreasing density.
But it is still useful for a comparison between the different models.
In the case without a planet, we assume that $\dot{m}_\mathrm{cross,in} = \dot{m}_\mathrm{visc}$, so that the mass accretion rate onto the star balances the resupply from the outer to the inner disc and the inner disc mass decreases with the wind mass-loss rate.
The inner disc would then be dispersed after only 80~kyr.
In the \textsc{pewind} model, the lifetime would be extended by more than a factor of 3 to 260~kyr.
In the \textsc{nowind} model, the inner disc would be accreted by the star within 455~kyr.
If the planetary gap acts as a filter for the dust in the disc, the increased lifetime of the inner gas disc could potentially explain the higher-than-expected occurrence rate of observed accreting transitions discs with large dust cavities \citep[e.g.][]{ercolano2017}.

The prolonged lifetime of the \textsc{pewind} model compared to a model with no planet is opposite to the expectation in the proposed planet-induced photoevaporation (PIPE) scenario \citep{alexander2009, rosotti2013}, which is based on the assumption that a gap-opening planet reduces the rate of viscous mass-transport to the inner disc.
Hydrodynamic simulations show that planets typically reduce the mass flow across the gap to 10 -- 25 per cent of the viscous steady-state accretion rate \citep{lubow2006}.
Considering the wind recycling and the additional mass-loading across the gap, PIPE can only operate if this reduction exceeds the effects of the wind-driven redistribution, which could be the case for more viscous discs with higher accretion rates.

The effects of the gap on the photoevapoative wind depend primarily on the width of the gap, which increases with increasing orbital radius and decreasing planet-to-star mass ratio, disc scale height and viscosity \citep{kanagawa2015}.
We therefore expect the effects to be more substantial if the disc is less viscous, the planet more massive, or the orbit larger. 
However, the exact relation between the width of the gap and the size of the affected wind region is unclear because the latter also depends on the temperature and density gradients in the wind itself, which are not scale-invariant so that more models would have to be run to make accurate predictions.

\subsection{Observational signatures}
In a companion work \citet{weber2022} have studied the effect that the modified wind structure has on the spectral line profiles of forbidden emission lines that are commonly used as disc wind tracers (e.g. \ion{O}{i}~6300~\AA{}).
They found that while the region above the gap, where the kinematics of the wind is affected the most, is not very well traced by the wind, the asymmetric substructures that are generated by the planet do leave an imprint in the wind that can cause variations in the line profiles that could be observed with high spectral resolution ($R \approx 100\,000$).
Most notably, the peak of the line profiles can move between the blue- and the redshifted parts of the spectrum on timescales of less than a quarter of the planet's orbit.

In recent years, molecular line observations with ALMA have very successfully been used to detect kinematic signatures of planets via the distortion of the disc's gas flow (e.g. \citealp{pinte2018, teague2019, wolfer2023}; review by \citealp{pinte2022}). In particular, it has become possible to map the 3D gas flow in the vicinity of the planet and, by doing so, to reveal exciting signatures, for example, of meridional flows \citep[e.g.][]{teague2019,yu2021}. \citet{galloway-sprietsma2023} even found evidence for an upward flow arising from the location of the gap of AS 209, which could be evidence for a disc wind. In light of these new possibilities, investigating our photoevaporative wind's effect on these observations would be extremely interesting. Unfortunately, the spatial resolution of the observations currently limits this kind of analysis to planets at larger separations than that in our model. However, we will present in a future work synthetic observations of our model with a planet at a larger distance from the star to investigate the effect that photoevaporation has on the kinematic signatures of the planet. 

One of our main findings in this work is that the photoevaporative wind refills the gap, reducing its depth by more than an order of magnitude. This possibility should be considered when inferring disc or planet properties, such as the viscosity parameter $\alpha$ or the planet's mass from observed gap depths \citep[e.g.][]{kanagawa2015, liu2018}.

\subsection{Mixing and reprocessing of disc material}
Our results show that a photoevaporative wind can redistribute material from the inner disc towards the gap and the circumplanetary region, from where it can be accreted by the planet or accumulate near the meridional flows at the gap edges.
This could have significant consequences on the gas composition near the gap or in the planet's atmosphere because the gas in the inner disc is expected to have a different composition than that in the outer disc. For example, inward-drifting and evaporating pebbles can significantly enrich the inner disc with volatile vapour (e.g. H$_2$O inside the water snowline) \citep{booth2017, schneider2021}. 
If it is then transported through the wind towards the planet, this could enrich the planet's atmosphere with heavy elements, even if the planet is located beyond the respective snowlines of the volatiles.
Such enrichment is not only observed for most hot-Jupiters \citep{thorngren2016} but also for the giant planets in the solar system, which all have an enhanced heavy element mass-fraction compared to the solar value \citep[e.g. review by][]{guillot2022}.

It should also be considered that when gas or dust is transported through the wind or along the disc surface, it is subject to a high-temperature environment with highly energetic EUV or X-ray irradiation from the central star. 
This allows for a reprocessing of the material that is impossible in the disc's interior.
One such process is isotope selective photodissociation, an essential mechanism for isotope fractionation, especially for oxygen and nitrogen isotopes (\citealp{visser2009, furuya2018}; review by \citealp{nomura2022}). 
Isotopic ratios are valuable tools for tracing the chemical evolution and origin of objects in our solar system. However, to link them to planetary formation theories, it is necessary to take into account the environment that the gas and dust were already subject to in the protoplanetary disc. 

Besides the gas, photoevaporation could also help redistribute small dust grains in the disc. Using the same photoevaporation model as in our model, \citet{franz2020,franz2021} have shown that the photoevaporative wind can entrain dust grains with a size of up to $\approx 11~\mu m$.
This depends strongly on the amount of small grains that are available on the disc surface, thus on the degree of vertical mixing in the disc, for example, by the vertical shear instability (VSI) \citep[e.g.][]{flock2017,flock2020}.
In Sect. \ref{sec:res:redistribution}, we have also seen that the meridional flows, where some of the redistributed material accumulates, contribute significantly to mass delivery into the gap. They could also stir up dust grains from the midplane towards the surface, where they could undergo further reprocessing.

Addressing the above points will require the detailed post-processing of our model to calculate the chemical evolution of the gas and, ideally, the dust. While this is out of scope for this work, the model presented here can serve as the basis for future investigations.

\subsection{Comparison with magnetic wind models}
Several studies that have been published recently have investigated the interplay between planet-hosting discs and magnetic winds.
In particular, \citet{aoyama2023} and \citet{wafflard-fernandez2023} independently found that the magnetic field lines concentrate in the gap and are deflected by the planet such that a magnetic wind is driven most dominantly from the outer edge of the gap.
This is in contrast to our findings with a photoevaporative wind, which promotes a vertical infall into the gap also at the outer edge. 
As we find in our model, \citet{wafflard-fernandez2023} report reduced migration rates of the planet in the presence of an MHD wind, a result of their gap becoming asymmetric with the outer gap being deeper radially outside of the planet than inside.
\citet{aoyama2023} also found that the gaps become deeper and wider in their MHD simulations.

It is likely that, in reality, disc winds are not either purely magnetic or purely thermal, but both mechanisms operate simultaneously, and their relative importance may vary with the evolutionary stage of the disc \citep[e.g.][]{pascucci2020}. 
In that case, a magnetic wind could act against the infall of a thermal component in the wind if it is strong enough to overcome the inverted thermal pressure gradient and sustain the wind above the gap.
However, if the magnetic wind cannot completely prevent the infall, some of the effects observed in both models could amplify each other.
For instance, the reduced migration rate is observed with both types of winds. It is also expected that with wider gaps, a larger volume of the wind is affected.
Wind models that combine both (non-ideal) magnetohydrodynamic effects and photoevaporation currently exist only for 2D axisymmetric discs without planets, but they still provide very useful insights into the wind-launching processes.

\citet{wang2019} developed non-ideal MHD disc wind models including self-consistent thermochemistry.
They found that their magnetothermal wind is predominantly launched by the magnetic toroidal pressure gradient and that radiative heating affects the wind launching only weakly, although the mass-loss rate of their wind depends on the ionisation fraction near the wind base and is thus moderately affected by the luminosity of high-energy radiation, mainly soft-FUV and Lyman-Werner photons.
In their model, the X-ray luminosity affects the accretion rate rather than the wind mass-loss rate.
However, they only include 3~keV X-ray photons, and as was shown by \citet{ercolano2009}, the heating in the wind-launching region is by far dominated by the soft X-ray band (< 1~keV).
In a similar model with more complete radiation physics (but still mainly focused on FUV-driven processes), \citet{gressel2020} found a wind that is launched magnetocentrifugally rather than by magnetic toroidal pressure, so the mass loss is also only moderately affected by thermochemical effects.
In both cases, the dynamics of the magnetic wind do not depend strongly on the thermal pressure, and its behaviour in the presence of a gap would depend on the impact of the gap on the magnetic field.

Other models indicate that thermal pressure may play an important role even in the presence of a magnetothermal wind:
\citet{rodenkirch2020} modelled a wind that includes ohmic diffusion as a non-ideal MHD effect and implements the same temperature and ionisation parametrisation by \citet{picogna2019} that is used in this work.
They found that photoevaporation is the dominant launching mechanism for weak magnetic fields with a plasma parameter $\beta > 10^7$.
With a different model that includes all non-ideal MHD effects and again the same temperature parametrisation, \citet{sarafidou2023} recently found that while the inner regions of the disc ($R \leq 1.5$~au) are always dominated by magnetic winds, photoevaporation can dominate the mass loss at larger radii for stronger magnetic fields ($\beta \geq 10^5$ or even stronger), depending on the X-ray luminosity.
Moreover, they found that beyond the footpoint of the wind, it is generally sustained thermally rather than magnetically.
In that case, the thermal pressure may dominate over the magnetic field in the region above the gap where the pressure gradient is inverted, such that the infall is sustained against the magnetic wind, especially from the inner disc, considering the magnetic wind is expected to be weaker at the inner edge of the gap.
However, without detailed models of non-ideal magnetothermal winds in protoplanetary discs with gaps, the outcome of this scenario remains unclear.

\section{Summary}\label{sec:summary}
We have performed three-dimensional radiation-hydrodynamic simulations of a photoevaporating, viscous ($\alpha = 6.9\cdot10^{-4}$) protoplanetary disc hosting a Jupiter-like planet.
We studied the interplay between the photoevaporative wind and the substructures generated by the planet, particularly the gap, and traced the redistribution of the gas and its accretion onto the planet. For comparison, we have also run the model with photoevaporation switched off. We refer to the model with and without photoevaporation with the identifiers \textsc{pewind} and \textsc{nowind}, respectively.
Our results show that:
\begin{itemize}
    \item At the location of the gap carved by the planet, a photoevaporative wind cannot be supported.
    In the vicinity of the gap, the pressure gradient in the wind is inverted, and the wind is falling back into the gap. Consequently, the depth of the gap is reduced by more than an order of magnitude.
    The same mechanism is at work not only in the wind but also at the hot, uppermost layers of the disc surface at the inner and outer edge of the gap.
    \item The continuous resupply of gas into the gap significantly increases the amount of mass the planet has available for accretion.
    In our simplified accretion model, this leads to a factor 2 increase in the planet's accretion rate.
    Most of the additionally accreted gas originates from a radius up to $\approx 8$~au, beyond which the wind structure is not significantly affected by the presence of the gap. 
    It is particularly noteworthy that $\approx 3$~percent of the accreted mass originates inside $R < 3$~au if photoevaporation is operating.
    In the \textsc{nowind} model, this region does not contribute to the accreted mass in a significant amount.
    \item The modified density structure of the disc leads to a reduction of the torque that is acting on the planet.
    After 200 orbits, the migration timescale is already reduced by almost a factor of 2 compared to the case without photoevaporation, and the torque is expected to continue to decrease.
    \item After being delivered to the gap, the material that is not accreted by the planet is pushed towards the edges of the gap by the gravitational torque of the planet, from where it viscously diffuses in the bound disc.
    This leads to an increase of the gap-crossing rates by a factor of $\approx 5$ in both directions compared to the \textsc{nowind} model.
    Consequently, the mass rate crossing the gap from the outer disc to the inner disc in the \textsc{pewind} model is higher than the viscous accretion rate that would be expected in an unperturbed disc without a planet.
    \item The increased inwards-crossing rate, in combination with the recycling of a significant portion of the wind that is falling back to the gap, results in a slower dispersal and, therefore longer lifetime of the inner disc compared to a model without a planet.
    This is opposite to the planet-induced photoevaporation (PIPE) scenario that has been proposed in the past.
\end{itemize}

We have discussed that the wind-driven redistribution of gas could potentially influence the chemical and isotopic composition of the planet's atmosphere and the material in the vicinity of the gap.
In future work, we will present an adapted model more suitable for creating synthetic observations of molecular (CO) emissions and compare them to actual observations.
It is unclear how a magnetic wind would change our results. Models of non-ideal magnetothermal winds in protoplanetary discs with gaps will be valuable for determining whether an infall would also happen in the presence of magnetic fields.


\begin{acknowledgements}\label{sec:acknowledgements}
      We would like to thank the anonymous referee for a constructive report that improved the manuscript. This research was supported by the Excellence Cluster ORIGINS, which is funded by the Deutsche Forschungsgemeinschaft (DFG, German Research Foundation) under Germany´s Excellence Strategy – EXC-2094-390783311. GP and BE acknowledge the support of the Deutsche Forschungsgemeinschaft (DFG, German Research Foundation) - 325594231. The simulations have been carried out on the computing facilities of the Computational Center for Particle and Astrophysics (C2PAP).
\end{acknowledgements}

\bibliographystyle{aa}
\bibliography{main}

\begin{appendix}
    \section{Influence of the diffuse EUV and X-ray field on the temperature structure of the gap}\label{sec:appendix-Tgap}
\begin{figure}[htbp!]
    \centering
    \resizebox{\hsize}{!}{\includegraphics{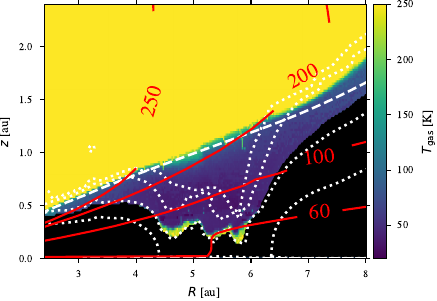}}
    \caption{
    Temperature structure of a density slice at $\Phi=0$ post-processed with \textsc{mocassin}.
    The white dashed line shows the contour for a column number density of $2.5\cdot10^{22}$~cm$^{-2}$, above which the temperature is fixed to the dust temperature.
    The red solid lines are contours of the dust temperature.
    The dotted white lines are density contours for (from top to bottom) $10^{-16}$ to $10^{-13}$~g\,cm$^{-3}$ in increments of one order of magnitude and indicate the structure of the gap.
    The black region is too far inside the disc to receive any energy packets in the radiative transfer calculation, which means it is shielded from stellar radiation or the diffuse field.
    }
    \label{fig:T_gap_diffuse}
\end{figure}

A shortcoming of the temperature parametrisation used in our models is that any heating by diffuse EUV or X-ray radiation is neglected.
To verify that this effect is not strong enough to affect our result significantly, we took a density slice of our model at $\Phi=0$ in the $R\theta$-plane and post-processed it with \textsc{mocassin}, using the same configuration as was used by \citet{picogna2019} to derive the parametrisation.
We present the resulting temperature structure inside the gap in Fig. \ref{fig:T_gap_diffuse}.
Most of the gap is not significantly heated by the diffuse field, and the temperature remains below the dust temperature.
In our model, the dust temperature is assumed where the column number density exceeds $2.5\cdot10^{22}~$cm$^{-2}$. Hence, the diffuse field has only an effect if it can heat the gas beyond the dust temperature.
This is the case only deep inside the gap, close to the 10$^{-14}$~g\,cm$^{-3}$ density contour, where the temperature reaches up to $\approx 250$~K.
However, this temperature and the resulting pressure gradient are more than an order of magnitude lower than the opposing gradient at the surface of the disc and not sufficient to launch a photoevaporative wind inside the gap. We can illustrate this using the basic concept of the gravitational radius \citep{liffman2003}, 
\begin{equation}
    R_g = \frac{\gamma-1}{2\gamma}\, \frac{G M_* \mu m_p}{k_B T},
\end{equation}
which is often used to approximate the radius outside which a photoevaporative wind can be launched.
Here, $m_p$ is the proton mass, $k_B$ is the Boltzmann constant, and T is the gas temperature.
For $T = 250$~K, we find $R_g \approx 100$~au, much larger than the radius of the gap.

\section{Snapshots of gas tracers}\label{sec:appendix-tracers}
\begin{figure*}[htbp!]
    \centering
    \includegraphics[width=17cm]{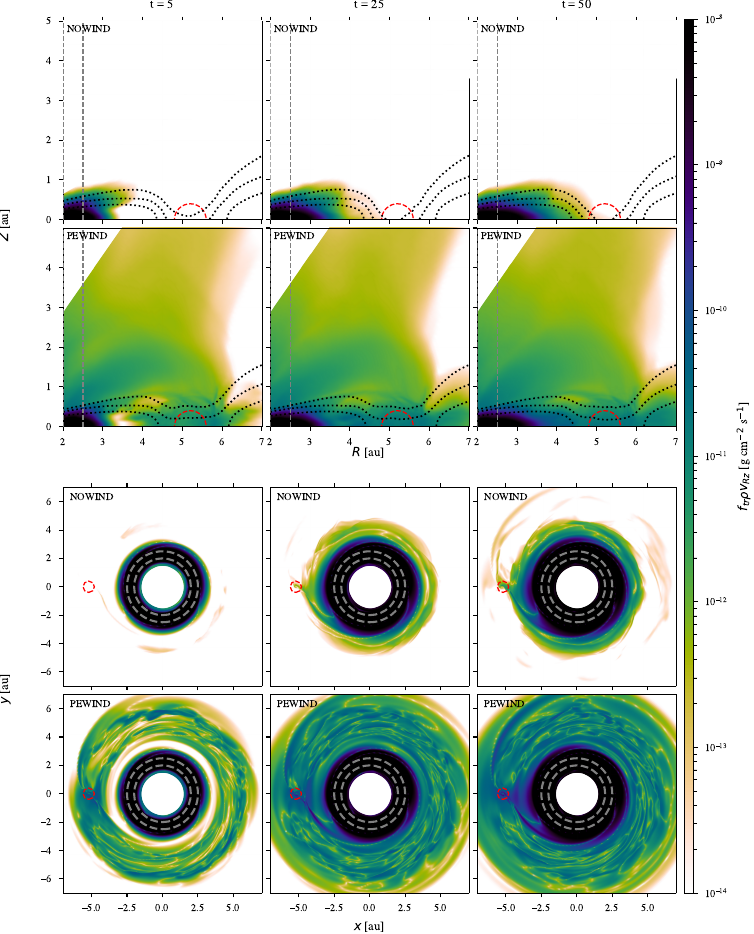}
    \caption{Poloidal mass flux of the gas originating in the innermost radial tracer-bin between $R = 2$ and 2.5~au at t = 5, 25 and 50 orbits. The top panels show the azimuthal average, and the bottom panels show the traced mass flux in the midplane. The black dotted lines are density contours for (from top to bottom) $10^{-17}$ to $10^{-13}$~g/cm$^3$ in increments of one order of magnitude. The red circle indicates the extent of the planet's Hill sphere. The grey dashed lines show the boundaries of the radial bin, inside which the tracer was originally distributed.}
    \label{fig:stracer-0_evol}
\end{figure*}

\begin{figure*}
    \centering
    \includegraphics[width=17cm]{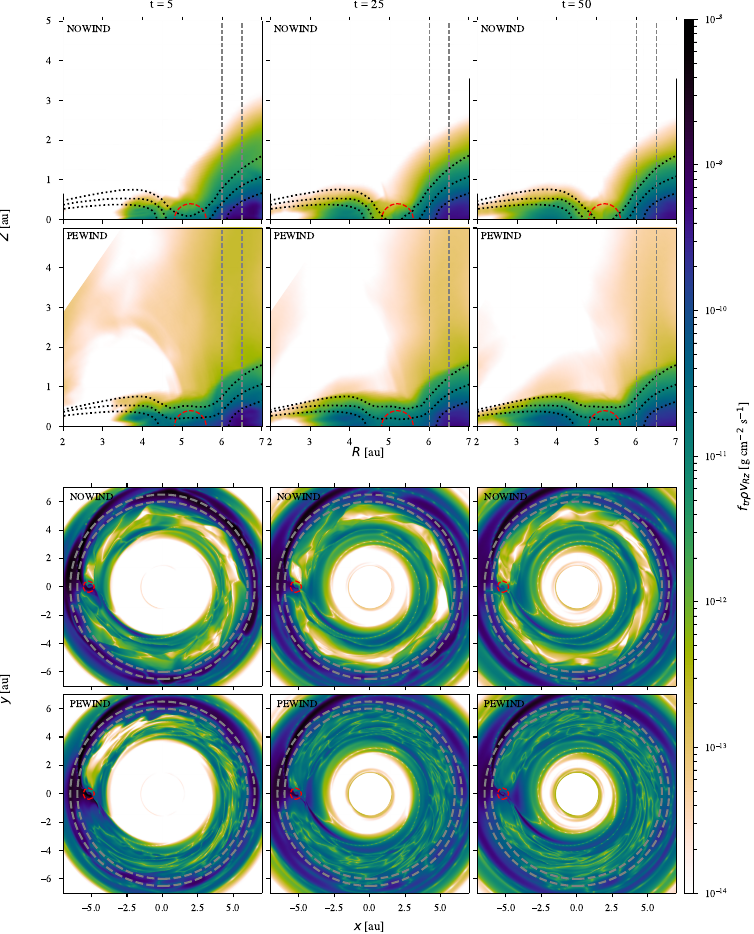}
    \caption{As Fig. \ref{fig:stracer-0_evol} but for the tracer that originates in the radial bin between $R = 6$ and 6.5~au}
    \label{fig:stracer-8_evol}
\end{figure*}
\end{appendix}

\end{document}